\documentclass[prd,aps,showpacs,secnumarabic]{revtex4}
\usepackage[dvips]{graphicx,color}
\usepackage{amssymb,amsmath,bm}
\def\al{\alpha}

\def\ga{\gamma}

\def\ep{\epsilon}
\def\la{\lambda}
\def\th{\theta}

\def\Dot{\!\cdot\!}

\def\Tr{{\rm Tr}}
\def\sl#1{#1\hspace{-5pt}/}
\def\Sl#1{#1\hspace{-7pt}/}
\def\Dot{\!\cdot\!}

\begin{document}
\title{An estimate of the branching ratio for ${\bm Z\bm\to\bm\nu\bar{\bm\nu}\bm\ga\bm\ga}$}
\author{Duane A. Dicus} \affiliation{Center for Particle Physics, University of Texas, Austin, TX 78712}\email{dicus@physics.utexas.edu} 
\author{Wayne W. Repko} \affiliation{Department of Physics and
Astronomy, Michigan State University, East Lansing, MI 48824}\email{repko@pa.msu.edu}
\date{\today}
\begin{abstract}
The effective interaction for two neutrino, two photon coupling is used to find an approximate width for the decay of the $Z$ boson into the $\nu\bar{\nu}\ga\ga$ final state.
\end{abstract}
\maketitle

It is somewhat surprising that an experimental upper bound exists for the branching ratio of $Z\to\nu\bar{\nu}\ga\ga$. In the early 1990s, both the L3 and OPEL groups looked for the decay at LEP. The net result was the limit \cite{LEP} $BR(Z\to\nu\bar{\nu}\ga\ga)\leq 3.1\times 10^{-6}$. Larios et al. \cite{LPT-V} used this result together with a model where the neutrinos have a magnetic moment to put limits on the magnetic moment of $\nu_\tau$ and, with a model of scalar couplings between the neutrinos and the photons, to put limits on the scalar couplings. (Of course, if the neutrinos are Majorana they do not have magnetic moments.)

It is also surprising that there exists an effective two neutrino-two photon coupling \cite{ADDR,DKR}
\begin{equation}
{\cal L}^{\rm SM}_{\rm eff}=\frac{1}{32\pi}\frac{ig^2\al}{M_W^4}A\left[ \bar{\psi}\ga^\nu(1-\ga^5)(\partial^\mu\psi)-(\partial^\mu\bar{\psi})\ga^\nu (1-\ga^5)\psi\right]F_{\mu\la}F^{\la}_\nu\,,\label{Leff}
\end{equation}
where $g$ is the electroweak gauge coupling, $\al$ is the fine structure constant, $\psi$ is the neutrino field, $F_{\mu\nu}$ is the electromagnetic field tensor.

For energies less than $m_e$ the parameter $A$ is given by \cite{ER,DKR}
\begin{equation}\label{A}
A=\left[\frac{4}{3}\ln\left(\frac{M_W^2}{m_e^2}\right)+1\right].
\end{equation}
For the energies of interest here $A$ is given by using Eq.\,(\ref{Leff}) to calculate $\gamma\gamma\,\rightarrow\,\nu\bar{\nu}$
and fitting to the numerical result shown in Fig. 4 of reference \cite{ADDR}.  This gives
\begin{equation}\label{AA}
A\,=\,13.66\,,
\end{equation}
independent of neutrino flavor, for center of mass energies from slightly above the charged lepton mass to $100$ GeV.

We can get an approximate expression for the amplitude for the $Z$ decay into two neutrinos and two photons by coupling the neutrino field, or the anti-neutrino field, of (\ref{Leff}) to the standard model $Z\nu\bar{\nu}$ interaction. This is obviously gauge invariant for the photons. Furthermore, if both neutrinos in the coupling (\ref{Leff}) are contracted, we obtain an amplitudes for $Z\to\ga\ga$ that must vanish if ${\cal L}^{SM}_{\rm eff}$ is a reasonable expression to use in $Z$ decay. This is not self evident, but can be seen from the resulting expression
\begin{equation}\label{Zgg}
T(Z\to\ga\ga)\sim\ep_\al(P)X_{\mu\nu}\int\!\!\frac{d^4s}{(2\pi)^4}\frac{(2s+P)^\mu} {s^2(s+P)^2}\Tr\left[\ga^\al\sl{s}\ga^\nu(\sl{s}+\Sl{P})(1+\ga^5)\right]\,,
\end{equation}
where $\ep_\al(P)$ is the polarization vector of the $Z$ with momentum $P$ and
\begin{equation} \label{X}
X_{\mu\nu}\equiv\langle\,k_1,\ep_1,k_2,\ep_2|F_{\mu\la}(0)F^\la_\nu(0)|0\rangle\,,
\end{equation}
is the matrix element of the photons. The integral can only depend on some combination of $P^\al P^\mu P^\nu$, $P^\al g^{\mu\nu}$, $P^\mu g^{\al\nu}$, and $P^\nu g^{\al\mu}$. The first two vanish since $\ep(P)\Dot P=0$. The final two are also zero for the same reason once we use $P^\mu X_{\mu\nu}\sim P_\nu$ or $P^\nu X_{\mu\nu}\sim P_\mu$ and $P=k_1+k_2$.

\begin{figure}[h]\centering
\includegraphics[width=2.1in]{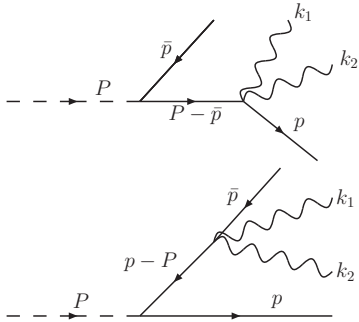}
\caption{\footnotesize The diagrams for $Z\to\nu\bar{\nu}\ga\ga$ using the $\nu\bar{\nu}\ga\ga$ effective Lagrangian are shown.  \label{Znngg}}
\end{figure}
Coupling (\ref{Leff}) to the $Z\nu\bar{\nu}$ vertex, as shown in Fig.\,(\ref{Znngg}), gives an amplitude for $Z\to\nu\bar{\nu}\ga\ga$ that has no free parameters. It is given by
%\newpage
\begin{equation} \label{M}
{\cal M} = \frac{g^3\al A}{64\pi\cos\th_W M^4_W}\bar{u}(p)
\left[\sl{\ep}\frac{(\sl{p}-\Sl{P})}{(p-P)^2}(\bar{p}-p+P)^\mu \ga^\nu 
+(\bar{p}-p-P)^\mu \ga^\nu \frac{(\Sl{P}-\sl{\bar{p}})} {(P-\bar{p})^2}\sl{\ep} \right](1-\ga^5) v(\bar{p})X_{\mu\nu}\,.
\end{equation}
After squaring and integrating over the four-body phase space, the resulting width for each neutrino is \cite{const}
\begin{eqnarray}
%\label{Gam}
\Gamma(Z\to\nu\bar{\nu}\ga\ga)&=&\frac{\alpha^5\,A^2}{\sin^6\th_W\cos^2\th_W}\frac{M_Z^9}{M_W^8}\frac{1}{90\,\pi^4}\frac{1}{2^{18}}
\left(\frac{28}{15}\pi^2-\frac{3146341}{173250}\right)\,,\label{Gamany} \\
                              &=&\,1.55\times 10^{-14}\,{\rm GeV} \label{Gam}\,,
\end{eqnarray}
or a branching ratio of $6.2\times 10^{-15}$.

We can also get an estimate of this partial width by calculating $Z\,\rightarrow\,Z^*H^*$, $Z^*\,\rightarrow\,\nu\bar{\nu}$, $H^*\,\rightarrow\,\gamma\gamma$ where $Z^*$ and $H^*$ are off-shell $Z$ or Higgs.  Using the expression for Higgs decay from \cite{HHG} or \cite{BP} with the Higgs mass replaced by the product of the photon momentum, $2k_1\Dot k_2$, we get
\begin{equation}\label{ZH}
\Gamma(Z\rightarrow\nu\bar{\nu}\gamma\gamma)\,=\,1.7\times10^{-17}\,{\rm GeV}\,.
\end{equation}
This is much smaller than the estimates above because it doesn't have the `box logarithm' factor $A$ of (\ref{Leff}).

\begin{figure}[h]
\centering
\includegraphics[width=1.75in]{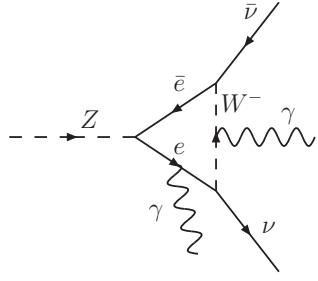}
\caption{\footnotesize One of the remaining diagrams contributing to $Z\to\nu\bar{\nu}\ga\ga$ is illustrated.  \label{Zee_nngg}}
\end{figure}

Of course, (\ref{Gamany}) or (\ref{Gam}) is only a rough estimate. In addition to the diagrams we have calculated, there are numerous other loop diagrams where the $Z$ decays into $e\bar{e}$ or $W^+W^-$ and the charged particles convert into neutrinos by the exchange of $W$ or an electron. The final photons are radiated from any of the charged particles in the loop. One example of such a process is shown in Fig.\,(\ref{Zee_nngg}).
But all these contributions are unlikely to give a total branching ratio larger than (\ref{Gam}) by more than an order of magnitude. What (\ref{Gam}) shows is that the branching ratio for this process is so small as to be not worthy of further calculation because it is unlikely to ever be measured.

\begin{acknowledgements}
DAD was supported in part by the U.~S.~Department of Energy under grant No. DE-FG02-12ER41830. WWR was supported in part by the National Science Foundation under Grant PHY 1068020.
\end{acknowledgements}


\begin{thebibliography}{99}
\bibitem{const} Since $\al$, $g$ and $\sin\th_W$ are related, we rewrote the coefficient of (\ref{M}) as $\al^{5/2}\sqrt{\pi}A/(8\sin^3\th_W\cos\th_WM^4_W)$ and used $\al=1/128.1$ and $\sin^2\th_W=0.231$ when evaluating the numerical result.
\end{thebibliography}
\end{document}